\documentclass[12pt]{article}
\usepackage{fullpage}
\usepackage{graphicx}
\newcommand{\be}{\begin{equation}}
\newcommand{\ee}{\end{equation}}
\newcommand{\bphi}{\mbox{\boldmath $\phi$}}

\font\mybb=msbm10 at 11pt

\def\bb#1{\hbox{\mybb#1}}

\newcommand{\news}{\setcounter{equation}{0}\quad}
\def\ben{\begin{equation}}
\def\een{\end{equation}}
\def\bea{\begin{eqnarray}}
\def\eea{\end{eqnarray}}
\begin{document}
\title{
\vskip 2cm Hopf solitons in the Nicole model}
\author{
Mike Gillard and Paul Sutcliffe\\[10pt]
{\em \normalsize Department of Mathematical Sciences,
Durham University, Durham DH1 3LE, U.K.}\\[10pt]
{\normalsize Email: 
\quad mike.gillard@durham.ac.uk,
 \quad p.m.sutcliffe@durham.ac.uk}
}
\date{July 2010}
\maketitle
\begin{abstract}
The Nicole model is a conformal field theory in three-dimensional
space. It has topological soliton solutions classified by the 
integer-valued Hopf charge, and all currently known solitons are axially 
symmetric. A volume-preserving flow is used to numerically construct
soliton solutions for all Hopf charges from one to eight. It is found
that the known axially symmetric solutions are unstable for Hopf charges
greater than two and new lower energy solutions are obtained that
include knots and links. A comparison with the Skyrme-Faddeev
model suggests many universal features, though there are some differences
in the link types obtained in the two theories.
\end{abstract}

\newpage
\section{Introduction}\news
Hopf solitons arise in theories in three-dimensional space
where the field takes values in a two-sphere. 
The Skyrme-Faddeev model \cite{Fa2}
is the most famous example of a theory with Hopf solitons and 
consists of the $O(3)$ sigma model modified by the addition
of a Skyrme term that is quartic in the derivatives of the field.
Substantial numerical work \cite{FN,GH,BS5,HS,Su} 
has led to a reasonable understanding of the minimal energy solitons in this
theory. There are axially symmetric solitons for all Hopf charges but
they are stable only for charges one, two and four. For all other 
charges the minimal energy solitons are less symmetric and include knotted
and linked configurations. It is unknown whether the appearance of 
knots and links as minimal energy solutions is a universal feature 
of Hopf solitons, since the Skyrme-Faddeev model is currently the only
theory in which non-axial Hopf solitons have been investigated. 


The Nicole model \cite{Ni} is a conformal field theory with 
Hopf soliton solutions. The conformal symmetry allows the consistent 
use of an axially symmetric ansatz in toroidal coordinates that reduces
the partial differential equation of the static theory to 
a single ordinary differential equation for a profile function \cite{BF}.
The toroidal ansatz involves a pair of integers $(n,m)$ 
associated with angular windings around the
two generating circles of the torus, and the corresponding Hopf charge is 
$Q=mn.$  A field configuration of this type will be denoted by 
${\cal A}_{n,m}$ and may be thought of as a two-dimensional baby Skyrmion
\cite{PSZ} with winding number $m$ embedded in the normal slice to a circle in
three-dimensional space, with a phase that rotates through an angle
$2\pi n$ as it travels around the circle once. 
The profile function and energy within the toroidal ansatz have been 
obtained \cite{Ni} in closed form when $Q=1$ and have been computed 
numerically \cite{ASVW} for a range of pairs $(n,m)$. 
In this paper we restrict to the case $Q>0,$ since
the situation for $Q<0$ is simply obtained by reversing an orientation.

An upper bound has been derived \cite{ASVW} for the minimal energy soliton 
$E\le cQ^\frac{3}{4},$ where $c$ is a known constant. This upper bound
takes the same form as in the Skyrme-Faddeev model \cite{LY}.
Furthermore, in the Nicole model it has been shown that the
energy  $E_{\rm axial}$ of the axial ansatz obeys a linear lower bound 
$E_{\rm axial}\ge \widetilde c Q,$ with a known constant $\widetilde c.$  
Together these results imply that for sufficiently large
Hopf charge the minimal energy Hopf soliton is not obtained from the 
axially symmetric toroidal ansatz. This suggests that Hopf solitons 
in the Nicole model may be similar to those in the Skyrme-Faddeev model,
where unstable axially symmetric solitons signal the appearance of
more exotic knotted and linked solutions. The purpose of the present
paper is to investigate this issue using numerical simulations of
the full nonlinear field theory, without any restrictions to axial
symmetry. This requires a numerical method that can overcome known
technical difficulties associated with simulations of a scale invariant
field theory. We introduce a method of volume-preserving
flow to deal with this issue and present the results of
this approach for all Hopf charges from one to eight.
 A comparison with the Skyrme-Faddeev
model suggests many universal features, including the appearance of
knotted and linked configurations, although there are some 
minor differences in the details of particular links that appear
at given Hopf charges.
\newpage 

\section{The Nicole model and volume-preserving flow}\news
The Nicole model \cite{Ni} is a rather exotic 
modification of the $O(3)$ sigma model, although it involves the 
same field $\bphi:\bb{R}^3\mapsto S^2,$ which is realized as a 
real three-component
vector $\bphi=(\phi_1,\phi_2,\phi_3),$ of unit length,
$\bphi\cdot\bphi=1.$ 
As we are concerned only with static solutions then the Nicole model
may be defined by its static energy
\be
E=\frac{1}{32\pi^2\sqrt{2}}\int (\partial_i\bphi\cdot\partial_i\bphi)
^\frac{3}{2}\ d^3x, \label{energy_ni}
\ee
where the normalization is chosen for later convenience.

The fractional power in (\ref{energy_ni}) is a novel modification
of the sigma model energy density and is clearly engineered to produce a 
conformal theory in three spatial dimensions. 
Although there is no known physical motivation for this theory,
there is a mathematical stimulus to investigate the model as
the conformal symmetry leads to interesting mathematical properties:
including an explicit exact solution for the $Q=1$ Hopf soliton.
By studying this theory it is possible to make comparisons with
Hopf solitons of the Skyrme-Faddeev model and hence determine
which features appear to be generic. This should help in 
understanding the properties of Hopf solitons in theories 
constructed from more conventional terms. 

 Finite energy boundary conditions require that the field tends to a
constant value at spatial infinity, which is chosen to be
$\bphi=(0,0,1)\equiv{\bf e}_3.$  This boundary condition
compactifies space to $S^3,$ so that the field becomes a map $\bphi:
S^3\mapsto S^2.$ 
There is a homotopy classification of such maps, given by the Hopf charge 
$Q\in\bb{Z}=\pi_3(S^2).$ This integer has a geometrical interpretation
as the linking number of two curves obtained as the preimages
of any two distinct points on the target two-sphere.

The position of the soliton is the closed curve obtained as the preimage of the
point $\bphi=(0,0,-1),$ which is
antipodal to the vacuum value ${\bf e}_3$ on the target two-sphere.
Hopf solitons are therefore novel string-like topological solitons.

The static field equation that follows from the variation of the energy
(\ref{energy_ni}) is the 
nonlinear partial differential equation
\be
\partial_i\partial_i\bphi
+(\partial_i\bphi\cdot\partial_i\bphi)\bphi
+\frac{(\partial_i\partial_j\bphi\cdot\partial_j\bphi)\partial_i\bphi}
{\partial_k\bphi\cdot\partial_k\bphi}={\bf 0}.
\label{eom_ni}
\ee

Using combinations of stereographic projection and the standard 
Hopf map, the solution with Hopf charge $Q=1$ is explicitly 
given by \cite{Ni}
\be
\frac{\phi_1+i\phi_2}{1+\phi_3}
=\frac{2\lambda(x_1+ix_2)}{r^2-\lambda^2+2i\lambda x_3}.
\label{Q1}
\ee
Here $r^2=x_ix_i$ and $\lambda$ is an arbitrary positive real constant
associated with the scale of the soliton. It is easily seen from
(\ref{Q1}) that the position of the soliton is the circle in the plane
$x_3=0$ with centre the origin and radius $\lambda.$ The fact that
$\lambda$ is arbitrary is a result of the conformal symmetry of the
Nicole model.

The energy of this soliton is $E=1,$ as a result of the convenient
normalization of the energy in (\ref{energy_ni}). The soliton
is axially symmetric and is of the type ${\cal A}_{1,1}.$

In this paper we are interested in Hopf solitons of the Nicole
model with $Q>1,$ for which explicit closed form solutions are
not available. Numerical methods are therefore required to find
solutions of equation (\ref{eom_ni}) that correspond to
local minima of the energy (\ref{energy_ni}), including the
global energy minimum in each sector with Hopf charge $Q$ that
we consider.

The discretization (and restriction to a finite simulation region)
involved in the numerical simulation of a conformal field theory 
breaks the conformal invariance of the continuum theory and
leads to technical difficulties. This has been studied in detail 
for solitons in a simpler conformal field theory, namely
the $O(3)$ sigma model in two-dimensional space.
The numerical discretization is equivalent to studying a lattice
version of the theory and in the planar case the resulting energy 
minimization leads to an exceptional configuration on the lattice \cite{Lu}.
In the continuum theory this is associated with a reduction of 
the soliton scale until it is of the same order as the lattice spacing,
at which point the topology of the soliton is lost as it unwinds
by essentially falling through the lattice \cite{LPZ}. 
In the case of the planar sigma model a novel lattice formulation
has been devised \cite{Ward} that preserves topology on the lattice.
However, the topology of Hopf solitons in the Nicole model
is significantly more complicated than that of solitons in the planar sigma 
model, and no analogous topology preserving lattice formulation is known.

As expected, a standard discretization and energy minimization of the Nicole model
leads to the same technical difficulties as described above. Namely, any
initial condition with $Q\ne 0$ shrinks until the size of the configuration is 
of the same order as the lattice spacing, upon which the topology is lost, leading
to a trivial vacuum solution with $Q=0$.
An additional, though related, issue is that in numerical simulations
Euclidean space is typically replaced by a finite region $\Omega,$
with the field fixed to its vacuum value on the boundary of $\Omega.$
This will be a good approximation if $\Omega$ is large compared
to the soliton size. However, in a conformal theory there is no fixed soliton
size and the restriction to a finite volume also results in the shrinking 
of a soliton. 
To overcome these difficulties we introduce a volume-preserving flow,
using ideas based on a similar approach developed in the context
of domain walls \cite{GS}.

To describe the construction of our volume-preserving flow 
we initially concentrate on the continuum theory defined in
a finite region $\Omega,$ with vacuum boundary conditions,
 $\bphi={\bf e}_3$ on $\partial\Omega.$ In the later numerical
simulations $\Omega$ will be taken to be a cube.

A standard method to minimize the energy (\ref{energy_ni})
is to evolve any given initial condition using gradient flow.
A flow which is proportional to gradient flow is given by
\be
\partial_{t}\bphi={\bf F},
\label{gf0}
\ee
where the force ${\bf F}$ is the left-hand-side of the static 
field equation (\ref{eom_ni}), and is proportional to the variation
of the energy (taking into account the constraint that $\bphi$
lies on the unit two-sphere). Theoretically, the end-point of this flow 
yields static solutions that solve equation (\ref{eom_ni}).
As described above, a numerical discretization introduces
a spatial lattice that breaks the conformal symmetry of the theory, so that
a numerical solution of a standard discrete version of (\ref{gf0}) results
in a Hopf soliton that continually shrinks during the flow.
A minimal energy Hopf soliton in the continuum theory 
has zero modes (in particular a scale invariance) 
associated with the conformal symmetry and the spatial lattice 
produces negative modes associated with the broken zero modes.
The idea is to modify the standard gradient flow (\ref{gf0})
by projecting out the component in the direction of the zero mode.
The subsequent discretization will then not produce negative modes
since their origin has been removed from the flow.  

Define the following volume associated with a field configuration
\be
V=\int_\Omega (1-\phi_3)\ d^3x.
\label{vol}
\ee
This volume will serve as a measure of the size of a soliton, since
the integrand is maximal along the position of the soliton and is
zero if the field takes its vacuum value. In particular, the shrinking of
a soliton that results from the numerical discretization corresponds to
a decrease of the volume $V$ during the gradient flow (\ref{gf0}).
The aim is to construct a modified version of gradient flow that
preserves the volume $V,$ and hence fixes the soliton scale.

Associated with the volume (\ref{vol}) is the gradient flow 
\be
\partial_{t}\bphi={\bf e_3}-\phi_3\bphi\equiv {\bf f},
\label{gf1}
\ee
which results in a decrease in the volume $V$ during this flow, since
the force ${\bf f}$ is proportional to the variation of $V.$
Define the inner product
\be
<{\bf f},{\bf g}>\ =\int_{\Omega} {\bf f}\cdot {\bf g}\ d^3x,
\label{inner}
\ee
then the volume-preserving flow is given by
\be
\partial_t\bphi={\bf F}-\frac{<{\bf F},{\bf f}>}{<{\bf f},{\bf f}>}{\bf f}.
\label{vpgf}
\ee
This flow has been constructed by taking the standard gradient flow 
and then projecting out the component due to the volume reducing 
flow (\ref{gf1}). The resulting flow is therefore orthogonal to the
volume reducing flow and hence preserves the volume. 
It is easy to prove that the volume-preserving flow (\ref{vpgf}) indeed
preserves the volume $V$ and reduces the energy $E$: the proof
is a simple modification of that presented in \cite{GS} but the
result should be obvious from the geometrical aspect of its construction.

Equation (\ref{vpgf}) is a nonlinear partial differential
equation but it is also nonlocal, because of the appearance of the inner
product. However, it can be solved numerically using standard finite
difference methods. In section \ref{sec-num} we shall present results 
for a scheme using fourth-order accurate finite difference approximations 
to spatial derivatives on a cubic lattice consisting of 
$151^3$ points with unit lattice spacing (taking advantage of the
scale invariance of the continuum problem in $\bb{R}^3$). 
The flow is evolved using a simple first-order accurate explicit method with
timestep $\Delta t=0.1.$ All inner products are evaluated by approximating
integrals by sums over lattice sites.

\section{Initial conditions}\news
Initial conditions, for a range of values of $Q,$ need to be provided 
for the volume-preserving flow algorithm discussed
in the previous section. We use the approach introduced in \cite{Su},
which is briefly reviewed in this section.

To construct an initial field the spatial coordinates 
$(x_1,x_2,x_3)\in \bb{R}^3$ are first mapped to the
unit three-sphere via a degree one spherically equivariant map.
Explicitly, introduce the complex coordinates $Z_1,Z_0$ (on the unit
three-sphere $|Z_1|^2+|Z_0|^2=1$) as
\be (Z_1,Z_0)=\bigg((x_1+ix_2)\frac{\sin g}{r},\cos
g+i\frac{\sin g}{r}x_3\bigg),\ee where 
the profile function $g(r)$ is a monotonically decreasing function of
the radius $r$, with boundary conditions $g(0)=\pi$ and
$g(\infty)=0.$ 

The initial condition is obtained by taking the stereographic
projection of $\bphi$ to be a rational function of $Z_1$ and $Z_0.$
The simplest example is to take 
\be
\frac{\phi_1+i\phi_2}{1+\phi_3}=\frac{Z_1}{Z_0},
\label{Qnm11}
\ee
which has $Q=1$ and is identical to the exact solution (\ref{Q1}) 
if the profile function is taken to be
 $g=\tan^{-1}(2r\lambda/(r^2-\lambda^2)).$

Other choices of rational functions and profile functions 
do not give exact solutions 
but do provide suitable initial conditions for the numerical simulation.

An obvious generalization of (\ref{Qnm11}) is given by 
\be
\frac{\phi_1+i\phi_2}{1+\phi_3}=\frac{Z_1^n}{Z_0^m}.
\label{Qnm}
\ee
This is an axially symmetric field with Hopf charge $Q=nm.$
As discussed later, under
volume-preserving flow this initial condition
yields the solution of type ${\cal A}_{n,m},$ obtained
previously using toroidal coordinates \cite{ASVW}.

Initial fields that are not axially symmetric and include knots and
links can be obtained from less symmetric rational maps.
For full details see \cite{Su} but some examples that are used in
this paper include the $(a,b)$-torus knot (here $a,b$ are 
coprime positive integers with $a>b$)
\be
\frac{\phi_1+i\phi_2}{1+\phi_3}=\frac{Z_1^\alpha Z_0^\beta}{Z_1^a+Z_0^b},
\label{Qab}
\ee
where $\alpha$ is a positive integer and $\beta$ is a non-negative 
integer. The Hopf charge of this field is $Q=\alpha b+\beta a.$
The position of this field is an $(a,b)$-torus knot and we denote
a field of this type by ${\cal K}_{a,b}.$ Of particular relevance
will be the simplest torus knot, the trefoil knot, which corresponds
to $(a,b)=(3,2)$ and can be obtained with $Q=7$ from the choice $\alpha=2$
and $\beta=1.$

In all the configurations discussed above, the position curve contains
only a single component. However, Hopf solitons also exist in which the 
position curve contains disconnected components that are linked.
In the simplest case there are just two components and such a linked 
configuration will be denoted by the type ${\cal L}_{p,q}^{\alpha,\beta},$
where $p$ and $q$ denote the Hopf charges of the two components if each 
is taken in isolation, and $\alpha$ and $\beta$ are the additional 
contributions to the Hopf charge due the linking of each of the
components with the other. The total Hopf charge is therefore
$Q=p+q+\alpha+\beta.$  
Initial fields of this type can be constructed using a rational map
in which the denominator is reducible. As an example, the field
\be
\frac{\phi_1+i\phi_2}{1+\phi_3}=\frac{Z_1^{n+1}}{Z_1^2-Z_0^2},
\label{Qlink}
\ee
is of the type ${\cal L}_{n,n}^{1,1}.$ It consists of two Hopf 
solitons that are each of the topological type ${\cal A}_{n,1}$ and 
are linked once to create a field with Hopf charge $Q=2n+2.$

The construction described in this section is easily applied to
the situation of a finite simulation region $\Omega$ by replacing
the profile function boundary condition $g(\infty)=0$ by the
condition that $g=0$ on the boundary $\partial \Omega.$ 
 
\section{Numerical results}\label{sec-num}\news
In this section we present the results
of numerical computations of Hopf solitons using the volume-preserving
flow algorithm, with initial conditions constructed as in the
previous section. The simulation region $\Omega$ consists of the
cube $|x_i|\le 75$ with spatial derivatives approximated by
fourth-order accurate finite differences using a lattice spacing $\Delta x=1.$
The flow is evolved using an explicit method with
timestep $\Delta t=0.1$ and is terminated once the energy has
stabilized at a constant value. 

In the continuum theory in Euclidean space the energy of a given solution
is invariant under a spatial rescaling. The simulation lattice and finite
region $\Omega$ mean that the energy is not independent of the volume $V,$
but for a reasonable range of $V$ we find that there is only a weak variation of the
energy. The results of varying $V$ suggest that our quoted energies 
should be accurate to around one percent. In situations where there are
two different solutions with the same value of $Q,$ then comparisons of
energies are made using similar volumes. 
Generally the profile function
$g(r)$ is taken to have a simple linear form, with a cutoff to ensure
that the boundary condition is satisfied. Varying the cutoff allows different
values of $V$ to be investigated, and the results are found to be
consistent within the errors mentioned above, of around one percent.

As a first test the ${\cal A}_{1,1}$ solution with $Q=1$ is
computed using initial conditions (\ref{Qnm11}) 
with a simple linear profile
function. The energy is 
calculated to be $E=1.000,$ which agrees with the exact 
solution to three decimal places.
This accuracy is beyond that expected in general. 
Figure~1.1 displays the position curve for this $Q=1$ soliton. 
For clarity, a tube around the position is displayed by plotting 
an isosurface where $\phi_3$ is slightly greater than $-1$.
The stability of this solution has been confirmed by 
applying perturbations and also by using initial
conditions in which a squashing perturbation is applied to break
 the axial symmetry of the ${\cal A}_{1,1}$ rational map constructed
field. Under the volume-preserving flow the axial symmetry
is restored and the same solution is recovered.

The situation is similar for the $Q=2$ solution of type
${\cal A}_{2,1},$ which is shown in Figure~1.2
and has been constructed using initial conditions (\ref{Qnm}) with
$(n,m)=(2,1).$
This solution is
found to be stable under perturbations that break the axial
symmetry and its energy is computed to be $E=1.794.$  
This is expected to be the same solution as found using
toroidal coordinates, though our energy is $3\%$ 
lower than that reported in \cite{ASVW}. This difference
is slightly more than that expected from studying variations of the
volume $V.$ A comparison of our results for axially symmetric
solutions with larger values of $Q$ reveals a better agreement
with the toroidal computations in \cite{ASVW}, so the origin of
the unusually large disagreement in the case of $Q=2$ remains
unclear. We have also computed a solution of the type 
${\cal A}_{1,2}$ and confirmed the results of \cite{ASVW}, that
this solution has an energy greater than that of the 
type ${\cal A}_{2,1}.$ This result agrees with the similar
situation in the Skyrme-Faddeev model.

In the Skyrme-Faddeev model the axially symmetric 
$Q=3$ solution of type ${\cal A}_{3,1}$ is unstable to
a perturbation that breaks the axial symmetry \cite{BS5} to
form a twisted ring. The notation $\widetilde {\cal A}_{3,1}$
is used to denote the type of this solution, to indicate that
it has the same topological type as ${\cal A}_{3,1}$ but that
the axial symmetry is broken.
We have confirmed that a similar situation occurs 
in the Nicole model by first constructing the ${\cal A}_{3,1}$
solution,
using initial conditions (\ref{Qnm}) with
$(n,m)=(3,1),$
 and then applying a non-axial perturbation. The resulting
twisted ring solution is displayed in Figure~1.3 and its energy
is listed in Table~\ref{tab-energy}, together with the
energies of all the minimal energy solutions found for 
$1\le Q\le 8.$

\begin{figure}
\begin{center}
\includegraphics[width=16cm]{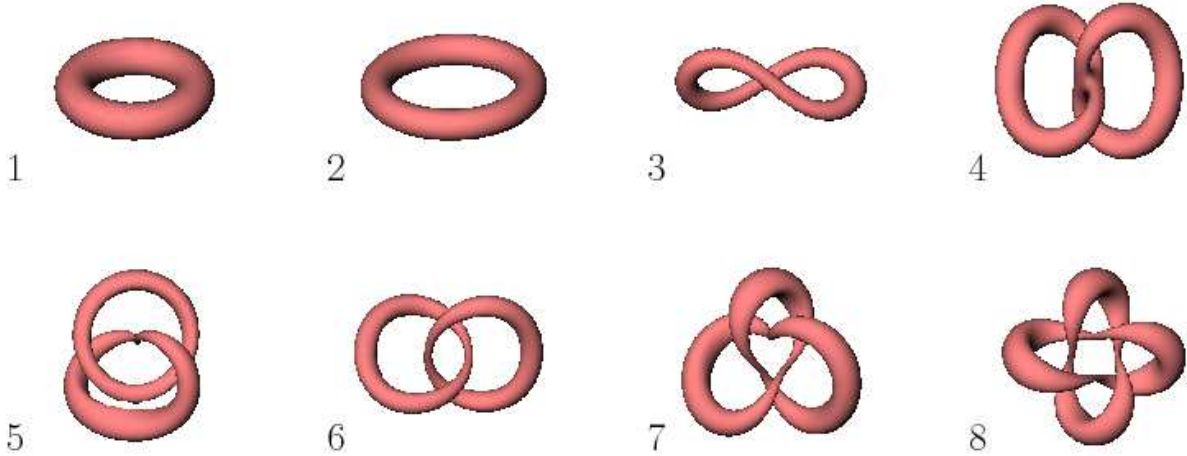}
\caption{Position curves for the known lowest energy solitons with Hopf
charges $1\le Q\le 8.$}
\label{fig-all8s}
\end{center}
\end{figure}

\begin{table}[ht]
\centering
\begin{tabular}{|c|c|c|c|}
\hline
$Q$ & Type & $E$ & $E/Q^{3/4}$ \\ \hline
1 & ${\cal A}_{1,1}$ & 1.000 & 1.000\\ 
2 & ${\cal A}_{2,1}$ & 1.794 & 1.067\\ 
3 & $\widetilde {\cal A}_{3,1}$ &2.535  &1.112 \\ 
4 & ${\cal L}_{1,1}^{1,1}$ & 3.156 & 1.116\\ 
5 & ${\cal L}_{1,2}^{1,1}$ &3.739  & 1.118\\ 
6 & ${\cal L}_{2,2}^{1,1}$ &4.339  & 1.132\\ 
7 & ${\cal K}_{3,2}$ &4.843  & 1.125\\ 
8 & ${\cal L}_{2,2}^{2,2}$ & 5.297 & 1.114\\ \hline
\end{tabular}
\caption{Energies and types 
of the known minimal energy solitons with $1\le Q\le 8.$
}
 \label{tab-energy}
\end{table}

\begin{figure}
\begin{center}
\includegraphics[width=10cm]{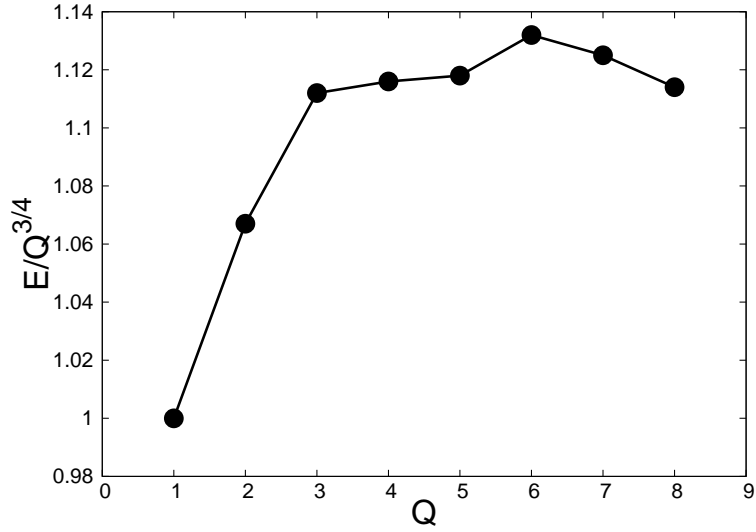}
\caption{The ratio of the energy $E$ to the conjectured bound
$Q^{3/4}$ for the known minimal energy solitons with $1\le Q\le 8.$ }
\label{fig-en}
\end{center}
\end{figure}

In the Skyrme-Faddeev model the minimal energy soliton
with $Q=4$ is axially symmetric and is of the type 
${\cal A}_{2,2}$ \cite{HS}.
A solution of this type has been constructed
in the Nicole model using toroidal coordinates \cite{ASVW}.
We have computed this solution,
using initial conditions (\ref{Qnm}) with
$(n,m)=(2,2),$
 but found that, in constrast 
to the Skyrme-Faddeev model, this solution is unstable
to perturbations that break the axial symmetry. Such a
perturbation leads to the linked solution displayed in
Figure~1.4, which is of the type ${\cal L}_{1,1}^{1,1}.$
We have also confirmed that the same solution is obtained
by starting with the initial conditions (\ref{Qlink}) with
$n=1.$ The energy of this solution can be found in 
Table~\ref{tab-energy}.

Solitons with $Q=5,6,7,8$ have been constructed using a 
variety of initial conditions for each $Q$, including perturbed axial
fields, links and knots. The resulting minimal energy solitons
for each $Q$ are displayed in Figure~1 and their energies and
types are listed in Table~\ref{tab-energy}. 
For $Q=5,6$ the solitons are links and have the same type as in
the Skyrme-Faddeev model. For $Q=7$ the soliton is a trefoil knot and
again this agrees with the result of the Skyrme-Faddeev model.
For $Q=8$ the soliton is a link of type  ${\cal L}_{2,2}^{2,2}.$
In the Skyrme-Faddeev model the minimal energy soliton with 
$Q=8$ is also a link,
but it has the different type ${\cal L}_{3,3}^{1,1}.$  In the
Skyrme-Faddeev model initial conditions of the type 
${\cal L}_{2,2}^{2,2}$ develop reconnections during the energy
minimization process and this produces the minimal energy link
${\cal L}_{3,3}^{1,1}$ \cite{Su}.

In the Skyrme-Faddeev model it has been shown \cite{VK} 
that a lower bound on the energy exists of the form 
$E\ge k\, Q^\frac{3}{4},$ where $k$ is a known
constant. To date, a similar lower bound has not been proved 
for the Nicole model: see \cite{ASVW} for a discussion of the 
technical difficulties in adapting the proof in \cite{VK} to 
the Nicole model. A conjectured lower bound for the Nicole model
is $E\ge Q^\frac{3}{4},$ where the constant has been set to 
unity, as this is the
largest possible value consistent with the energy of the explicit 
$Q=1$ solution. In the final column of Table~\ref{tab-energy}
we list the ratio of the energy to this conjectured bound and plot
this ratio in Figure~\ref{fig-en}. It is clear that our numerical
results are consistent with the conjectured bound. Furthermore, 
for $Q>2$ the excess above the bound appears to settle down to 
a value around $12\%,$ which suggests that the solutions we have
found are good candidates for the global energy minima for these
values of $Q.$

\section{Conclusion}\news
By introducing a volume-preserving flow we have been able
to numerically investigate Hopf solitons and their stability in the
Nicole model. It has been demonstrated that the known axially
symmetric Hopf solitons are unstable for Hopf charges greater
than two and new lower energy solutions have been computed
that include links and knots. The formation of links and knots
mirrors the situation in the Skyrme-Faddeev model, suggesting
that this is likely to be a universal feature of Hopf solitons.
However, for Hopf charges four and eight, links are formed 
in the Nicole model
that are not of the same topological type as in the Skyrme-Faddeev
model. As the Hopf charge increases there is an increased variety
of possible link and knot types, so it seems likely that it becomes
more common for the soliton types to disagree in the two theories.

A lower bound on the energy has been conjectured that is consistent
with the numerical results we have obtained. A proof of this 
conjectured bound would signal another universal feature of Hopf 
solitons.

The Aratyn-Ferreira-Zimerman (AFZ) model \cite{AFZ1} is another
conformal field theory with Hopf soliton solutions. 
As in the Nicole model, the conformal symmetry allows the consistent 
use of a toroidal ansatz to reduce to an ordinary differential 
equation for a profile function \cite{BF}.
In the AFZ model the profile function and associated energy 
can be obtained exactly in closed form \cite{AFZ2} 
for all solutions of the type ${\cal A}_{n,m}.$ 
This is a consequence of an additional infinite-dimensional 
symmetry group acting on target space. It would be interesting to
study the stability of these solutions and to investigate the 
existence of knotted and linked Hopf solitons in the AFZ model.
There are additional complications in the AFZ model, related to
the infinite-dimensional symmetry of the theory, as discussed
in \cite{ASW}. The approach of volume-preserving flow
therefore needs some modification to be applicable to this
model. This issue is currently under investigation.

\section*{Acknowledgements}
MG thanks the EPSRC for a research studentship.
PMS thanks the EPSRC for funding under grant number EP/G038775/1.
The numerical computations were performed on the Durham HPC cluster HAMILTON.

\end{document}